\documentclass{aa}
\usepackage{txfonts}
\usepackage{graphicx}
\usepackage{aalongtable}

\def\Lir{\hbox{$L_{\rm IR}$} }
\def\Sir{\hbox{$\Sigma_{\rm IR}$} }
\def\Rir{\hbox{$r_{\rm IR}^{}$} }
\def\Rirtwo{\hbox{$r_{\rm IR}^2$} }
\def\Air{\hbox{$\pi$\Rirtwo} }
\def\Thetarc{\hbox{$\theta_{\rm RC}^{}$}}
\def\Rrc{\hbox{$r_{\rm RC}^{}$} }

\def\Td{\hbox{$T_{\rm d}$} }
\def\Hii{\ion{H}{ii} }

\def\um{\hbox{\,$\mu$m} }
\def\Lsun{\hbox{\,$L_\odot$} }
\def\Lsunpc{\hbox{\,$\Lsun\,{\rm pc}^{-2}$} }
\def\Msun{\hbox{\,$M_\odot$} }
\def\K{\hbox{\,K} }
\def\kms{\hbox{\,km\,s$^{-1}$} }

\def\GHz{\hbox{\,GHz} }

\def\IRAS{{\it IRAS }}
\def\Spitzer{{\it Spitzer }}
\def\dx{\hbox{d}x}
\def\dt{\hbox{d}t}
\def\dnu{\hbox{d}\nu}

\begin{document}

   \title{The infrared compactness-temperature relation for quiescent and starburst galaxies}

   \titlerunning{The infrared compactness-temperature relation...}

   \author{P. Chanial\inst{1} 
          \and
        H. Flores\inst{2}
          \and
        B. Guiderdoni\inst{3}
          \and
        D. Elbaz\inst{5}
          \and
        F. Hammer\inst{2}
          \and
        L. Vigroux\inst{4,5}
          }

   \offprints{P. Chanial\\
              e-mail: {\tt p.chanial@imperial.ac.uk}}

   \institute{Astrophysics Group, Blackett Laboratory, Imperial College, Prince Consort Road, London SW7 2AZ, UK
         \and Laboratoire Galaxies, Etoiles, Physique et Instrumentation, Observatoire de Paris, 5 place Jules Janssen, F-92195 Meudon, France
         \and Centre de Recherche Astronomique de Lyon, Universit\'e Lyon 1, 
9 avenue Charles Andr\'e, F-69230 Saint-Genis Laval, France; CNRS, UMR 5574; Ecole Normale Sup\'erieure de Lyon, Lyon, France
         \and Institut d'Astrophysique de Paris, 98bis boulevard Arago, 75014 Paris, France; CNRS, UMR 7095; Universit\'e Pierre \& Marie Curie, Paris, France
         \and Service d'Astrophysique, DAPNIA, DSM, CEA-Saclay, Orme des Merisiers, B\^at. 709, F-91191 Gif-sur-Yvette, France
             }
   \date{Received July 22, 2005; accepted October 25, 2006}

   \abstract
   { \IRAS observations show the existence of a correlation between the infrared luminosity \Lir and dust temperature \Td in star-forming galaxies, in which larger \Lir leads to higher dust temperature. The $\Lir$--\Td relation is commonly seen as reflecting the increase in dust temperature in galaxies with higher star formation rate (SFR). Even though the correlation shows a significant amount of dispersion, a unique relation has been commonly used to construct spectral energy distributions (SEDs) of galaxies in distant universe studies, such as source number counting or photometric redshift determination.}
   {In this work, we introduce a new parameter, namely the size of the star-forming region \Rir and lay out the empirical and modelled relation between the global parameters \Lir, \Td and $\Rir$ of IR-bright non-AGN galaxies.}
   {\IRAS 60-to-100\um color is used as a proxy for the dust temperature and the 1.4\GHz radio contiuum (RC) emission for the infrared spatial distribution. The analysis has been carried out on two samples. The first one is made of the galaxies from the 60\um flux-limited \IRAS Revised Bright Galaxy Samples (RBGS) which have a reliable RC size estimate from the VLA follow-ups of the \IRAS Bright Galaxy Samples. The second is made of the sources from the 170\um ISOPHOT Serendipity Sky Survey (ISOSSS)  which are resolved by the NRAO VLA Sky Survey (NVSS) or by the Faint Images of the Radio Sky at Twenty-cm survey (FIRST).}
   {We show that the dispersion in the $\Lir$--\Td diagram can be reduced to a relation between the infrared surface brightness and the dust temperature, a relation that spans 5 orders of magnitude in surface brightness.}
   { We explored the physical processes giving rise to the $\Sir$--\Td relation, and show that it can be derived from the Schmidt law, which relates the star formation rate to the gas surface density.}
   \maketitle

%

\section{Introduction}

The sky survey by the \IRAS satellite (Neugebauer et al. \cite{IRAS}) led to the discovery of strong connections between global parameters of galaxies in the local universe.
Among them, the 60-to-100\um
flux density ratio $R$(60/100) versus \Lir \IRAS diagram (Soifer et al. \cite{BGS1a}) exhibits a large dispersion.
The quantities \Lir and $R$(60/100) of
quiescent and starburst galaxies are fundamental parameters for the study
of star formation.  The first one represents the energy absorbed and
reprocessed by dust and is related to the star formation rate.
The second parameter traces the dust temperature \Td and also provides an estimate of the star formation efficiency as defined by the SFR per unit of gas mass (Young et al. \cite{Young86}, Chini et al. \cite{Chini92}).

The first attempts to relate the infrared surface brightness (which we also refer to as infrared compactness) to the dust temperature  were
hampered by the lack of sufficient infrared spatial resolution.
Devereux (\cite{Devereux87}) used ground-based small-beam observations at 10\um and compared them to the large-beam 12\um \IRAS flux densities to estimate the compactness of optically bright galaxies. He showed that the ratio between the small-beam and large-beam flux densities is correlated with the global
\IRAS 12-to-25\um flux density ratio which also traces the dust temperature.
The main limitation of this work is the use of a rough compactness estimator that can not be easily related to physical parameters and thus no concluding relationship was derived.

Other similar attempts showed that the optical surface brightness (Phillips \& Disney \cite{Phillipps88}) and the infrared surface brightness derived from H$_\alpha$ effective area (Lehnert \& Heckman \cite{Lehnert96}) of IR bright galaxies increase with $R$(60/100). However, the quantitative understanding of these correlations is not straightforward because the optical surface brightness results from stars that may not be related to the dust emission and because in the second study, in addition to the small size of the sample (32 galaxies), the authors used H$_\alpha$ maps to estimate the star-forming region size without applying an extinction correction which turns out to be crucial (Chanial et al., in prep).

Wang \& Helou (\cite{Wang92}) made use of the extinction-free 1.4\,GHz radio continuum (RC) size estimators to show that the infrared luminosity is not proportional to the galaxy physical area. They also showed an empirical relation between the mean RC surface brightness $\Sigma_{\rm RC}$ and luminosity $L_{\rm RC}$, but they did not consider the IR color variations within their sample which, as we will show in this article, is directly related to the scatter in the $\Sigma_{\rm RC}$--$L_{\rm RC}$ relation.

More recently, Roussel et al. (\cite{Roussel01}) studied a  sample of galaxies mapped by the ISOCAM camera on board {\it ISO} (Cesarsky et al. \cite{ISOCAM}) and found a correlation between the 15-to-7\um flux density ratio and the 15\um effective surface brightness. Their sample only consists of quiescent spirals of moderate infrared luminosities and the authors only considered the circumnuclear region.

In this paper, we extend these approaches by (1) studying statistically larger samples, (2) applying a flux-limited selection criterion at 60$\um$, (3) considering fundamental parameters representative of the whole galaxy, and (4) using an extinction-independent estimator of the star-forming region size, the radio continuum angular size.

Section 2 describes the global parameters used in this paper and Sect. 3 presents the galaxy samples that are used in Sects. 4 and 5 for the analysis of the luminosity-temperature and compactness-temperature relations. The latter is modelled in Sect. 6 and discussed in Sect. 7.

\begin{table*}
   \caption{$R$(60/100) as an effective temperature estimator: the SCUBA 450\um SLUGS subsample.\label{table:slugs450}}
   $$
   \begin{array}{p{0.16\linewidth}p{0.07\linewidth}rcrrcc}
      \hline
      \hline
      \noalign{\smallskip}
\multicolumn{2}{l}{\rm Name \quad\quad\quad\quad\quad\quad Morphology} & D\;\; &\log \Lir& f_\nu(60\um)&f_\nu(100\um)&\Td(60/100)&\Td \\
      \noalign{\smallskip}
\quad(1)&(2)&(3)\;&\multicolumn{1}{c}{(4)}&\multicolumn{1}{c}{\quad\;(5)}&\multicolumn{1}{c}{\quad\quad(6)}&\multicolumn{1}{c}{(7)}&\multicolumn{1}{c}{(8)}\\
      \noalign{\smallskip}
      \hline
      \noalign{\smallskip}
UGC 903             & Sc   &  33.16 & 10.32 &  7.78 & 15.45 & 35.5\pm2.7 & 33.5\pm1.0 \\
NGC 958             & SBc  &  76.36 & 11.00 &  5.85 & 15.08 & 31.5\pm2.2 & 30.1\pm0.7 \\
UGC 2369\,S         & Sbc  & 121.90 & 11.39 &  8.07 & 11.18 & 40.4\pm3.3 & 40.7\pm1.5 \\
UGC 2403            & SBa  &  53.78 & 10.68 &  7.72 & 12.06 & 38.3\pm3.1 & 36.2\pm1.2 \\
NGC 2856            & Sc   &  40.08 & 10.32 &  5.73 & 10.15 & 37.0\pm2.9 & 36.2\pm1.3 \\
NGC 2990            & Sc   &  47.18 & 10.43 &  5.16 &  9.61 & 35.5\pm2.7 & 34.2\pm1.1 \\
UGC 5376            & Sd   &  31.28 & 10.10 &  5.36 & 10.41 & 34.7\pm2.6 & 33.1\pm1.0 \\
Arp 148\,E          & Irr  & 143.20 & 11.45 &  6.38 & 10.30 & 38.9\pm3.1 & 38.2\pm1.4 \\
Zw 247.020          & Sa   & 107.70 & 11.16 &  6.01 &  8.47 & 40.8\pm3.4 & 43.9\pm2.0 \\
I Zw 107            & Pair & 168.70 & 11.70 &  9.02 & 10.00 & 46.4\pm4.2 & 44.6\pm1.9 \\
NGC 5962            & Sc   &  32.17 & 10.41 &  8.93 & 21.82 & 32.3\pm2.3 & 31.3\pm0.8 \\
NGC 6052            & Pair &  70.42 & 10.86 &  6.79 & 10.57 & 38.3\pm3.1 & 36.8\pm1.3 \\
NGC 6181            & SABc &  30.70 & 10.36 &  8.94 & 20.83 & 32.7\pm2.4 & 33.3\pm1.0 \\
NGC 7541            & SBbc &  30.09 & 10.66 & 20.08 & 41.87 & 34.5\pm2.6 & 34.4\pm1.0 \\
NGC 520\,S          & Sa   &  30.22 & 10.78 & 31.52 & 47.37 & 39.1\pm3.2 & 40.7\pm1.3 \\
UGC 2982            & SABc &  67.57 & 10.97 &  8.39 & 16.82 & 34.7\pm2.6 & 35.4\pm1.0 \\
NGC 2623            & Sa   &  77.43 & 11.44 & 23.74 & 25.88 & 46.2\pm4.2 & 49.1\pm1.8 \\
NGC 3110            & SBb  &  73.48 & 11.17 & 11.28 & 22.27 & 34.7\pm2.6 & 36.7\pm1.1 \\
IRAS 10173+0828     &      & 198.70 & 11.63 &  5.61 &  5.86 & 49.5\pm4.6 & 46.9\pm1.7 \\
IRAS 10565+2448     &      & 176.30 & 11.87 & 12.10 & 15.01 & 43.2\pm3.7 & 47.2\pm1.6 \\
IRAS 12112+0305\,NE &      & 292.50 & 12.14 &  8.18 &  9.46 & 48.1\pm4.3 & 48.0\pm1.8 \\
UGC 8387            & Irr  &  99.99 & 11.55 & 17.04 & 24.38 & 36.2\pm2.8 & 41.3\pm1.2 \\
Zw 049.057          & Irr  &  59.06 & 11.20 & 21.89 & 31.53 & 40.0\pm3.3 & 40.9\pm1.3 \\
NGC 7592            & Pair &  95.13 & 11.17 &  8.05 & 10.58 & 42.0\pm3.5 & 38.7\pm1.5 \\
NGC 7714            & Sb   &  38.16 & 10.50 & 11.16 & 12.26 & 44.8\pm4.0 & 43.2\pm1.8 \\
      \noalign{\smallskip}
      \hline
   \end{array}
   $$
{\scriptsize Columns: (3) distance in Mpc; (4) infrared luminosity in $\Lsun$ derived from the 60 and 100\um \IRAS flux densities by assuming a modified blackbody of emissivity index $\beta$=1.3; (6) \& (7) \IRAS flux densities in Jy; (7) effective dust temperature and $1\sigma$ uncertainties in K, $\beta$=1.3; (8) effective dust temperature and $1\sigma$ uncertainties in K derived from the additional data points tabulated in Dunne \& Eales (\cite{Dunne01}).}
\end{table*}

\section{Parameter definitions and estimations}
\label{sect:params}
\subsection{The dust temperature}
\label{sect:dust-temperature}
The temperature of the dust in a given galaxy is subject to spatial variations (Dale et al. \cite{Dale99}) that show a decrease from the center outwards. So, the attempt to describe the global thermal emission of the dust with one or two modified blackbodies may be questioned.
However, submillimeter observations at 450\um and 850\um (Dunne et al. \cite{Dunne00}, Dunne \& Eales \cite{Dunne01}, Vlahakis et al. \cite{Vlahakis05}) have shown that the {\it global} emission at these wavelengths, which trace cold dust, does correlate with the global $R$(60/100), which traces warmer dust. This finding suggests that inner and outer emissions may be bound together.  Such a property would likely be the manifestation of the star formation regulation occurring on a global scale, as it appears in the Schmidt law (\cite{Schmidt59}) or in relations which involve the galaxy rotation curve such as the 
Toomre's stability criterion (\cite{Toomre64}) and the law by Kennicutt (\cite{Kennicutt98}).

\begin{figure}
  \centering
  \resizebox{\hsize}{!}{
    \includegraphics{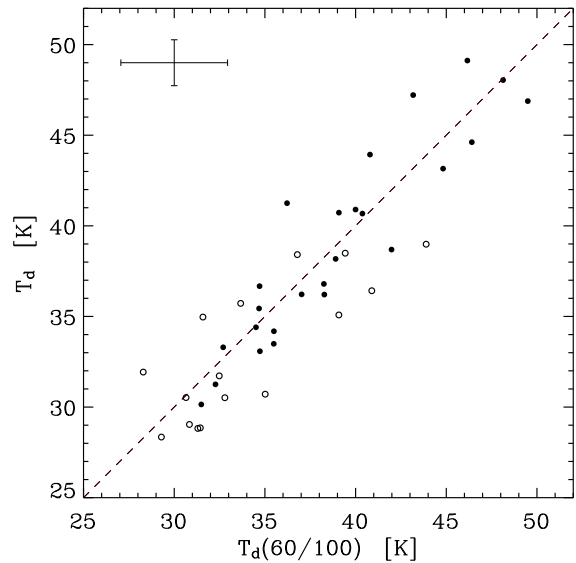}
  }
  \caption{ Temperature of the modified blackbodies ($\beta$=1.3) fitted to the 60 and 100\um \IRAS flux densities (absciss\ae) and to additional far-infrared and submillimeter observations (ordinates). The SLUGS 450\um subsample and the SINGS subsample are represented with filled and open circles. The dashed line is the unity line and the average uncertainties are shown in the upper left corner.}
  \label{fig:td-60-100}
\end{figure}

The often used two-component model does not reflect the fact that a single free parameter such as $R$(60/100) suffices to describe the SEDs of galaxies whose infrared emission arises from star formation (Dale et al. \cite{Dale01}, Dale \& Helou \cite{Dale02}). As a matter of fact, Serjeant \& Harrison (\cite{Serjeant05}) constructed a library of two-components IR templates, but to do so, they parametrized the temperatures and the relative weights with $R$(60/100).
By contrast, using a single blackbody with a constant emissivity index certainly does not provide the most accurate description of the SED, but it may supply a more profound insight on theglobal state of the dust in a galaxy.

We thus {\it define} for this paper the effective dust temperature \Td of a galaxy as the temperature of the modified blackbody of fixed emissivity index $\beta$ which best fits the galaxy rest-frame SED between 50 and 1000$\um$. The SCUBA Local Universe Galaxy Survey (Dunne et al. \cite{Dunne00}) is to date the largest homogeneous set of infrared bright galaxies observed at 850$\um$. The authors derived a mean emissivity index \hbox{$<$$\beta$$>$=1.3} with a standard deviation of 0.2. A subsample was then observed at 450\um (Dunne \& Eales \cite{Dunne01}) but the authors did not publish their single-temperature analysis, althought they do state that this model cannot be ruled out, and instead favoured the analysis of the two-component model which offers one additional free parameter. By excluding galaxies with an active galactic nucleus (AGN) from their 450\um sample and by taking the flux densities tabulated in their article, we found  a similar value of \hbox{$<$$\beta$$>$=1.38} with a sample standard deviation of 0.17. Since only 2 out of 25 galaxies have a reduced $\chi^2_r$$>$2, the single-temperature model provides a reasonable fit in most cases and we adopted the fiducial value $\beta$=1.3 in this paper.

The next step is to check that it is possible to easily associate an estimator to the previously defined effective dust temperature, the estimator of choice being the color $R$(60/100) due to its matchless availability in the local universe. We have already noted that the $R$(60/100) parameter can characterize to some extent the whole 
infrared SED of normal and starburst galaxies, but its precision still has to be determined.
For two sets of non-AGN galaxies described in Tables~\ref{table:slugs450} and \ref{table:sings}, we compared the effective dust temperatures exclusively derived from the 60 and 100\um \IRAS flux densities to the temperatures that are derived from additional far-infrared (FIR) and submillimeter observations. The two sets are the SCUBA SLUGS 450\um subsample (Dunne \& Eales \cite{Dunne01}) and the galaxies observed by \Spitzer at 70 and 160\um from the SINGS catalog (Kennicutt et al. \cite{SINGS}, Dale et al. \cite{Dale05}) for which submillimeter observations (800 or 850$\um$) were available in the literature. We completed the second set with far-infrared flux densities mainly from ISOPHOT (Lemke et al. \cite{ISOPHOT}) and the Kuiper Airborne Observatory.
The results are shown in Fig.~\ref{fig:td-60-100} and confirm $R(60/100)$ as an estimator of the effective dust temperature. The standard deviation is 2.8$\K$, which is small enough to validate the use of $R(60/100)$ to estimate the effective dust temperature in our study.


\begin{table*}
   \caption{$R$(60/100) as an effective temperature estimator:  the \Spitzer SINGS subsample.\label{table:sings}}
   $$
   \begin{array}{p{0.08\linewidth}p{0.05\linewidth}rrrrrr@{\quad\quad}lllcc}
      \hline
      \hline
      \noalign{\smallskip}
\multicolumn{2}{l}{\rm Name \quad Morphology} & \multicolumn{4}{l}{D \; \log \Lir \; f_\nu(60\um) \; f_\nu(70\um)} & \multicolumn{3}{l}{ f_\nu(100\um) \; f_\nu(160\um) \;\; \lambda}&\multicolumn{1}{l}{f_\nu(\lambda)}&{\rm Refs.}&\multicolumn{2}{l}{\Td(60/100)\quad\quad\Td} \\
      \noalign{\smallskip}
\quad(1)&(2)&\multicolumn{1}{l}{(3)}&\multicolumn{1}{l}{(4)}&\multicolumn{1}{c}{(5)}&\multicolumn{1}{c}{(6)}&\multicolumn{1}{c}{(7)}&\multicolumn{1}{r}{(8)\quad\quad}&\multicolumn{1}{l}{\quad(9)}&(10)&(11)&(12)&(13)\\
      \noalign{\smallskip}
      \hline
      \noalign{\smallskip}
NGC 337  &SBcd&21.6& 10.04 &  9.07&  8.83& 20.11& 18.30&850        &0.35            &1       &32.8\pm2.4&30.5\pm1.1 \\
NGC 2798 &SBa &27.8& 10.52 & 20.60& 14.70& 29.69& 18.45&850        &0.19            &1       &39.4\pm3.2&38.5\pm1.6\\
NGC 2976 &Sc  & 3.6&  8.68 & 13.09& 16.99& 33.43& 46.81&850        &0.61            &1       &30.8\pm2.2&29.0\pm1.2\\
NGC 3190 &Sa  &24.1&  9.80 &  3.33&  4.34&  9.84& 13.19&850        &0.19            &1       &29.3\pm2.0&28.3\pm0.9\\
Mrk 33   &Sm  &26.8&  9.83 &  4.79&  3.34&  5.49&  3.46&850,850    &0.05,0.04       &2,1     &43.9\pm3.9&39.0\pm1.5\\
NGC 3521 &SABb& 6.8&  9.81 & 49.19& 49.85&121.80&206.70&850        &2.11            &1       &31.3\pm2.2&28.8\pm1.2\\
NGC 4254 &Sc  &15.3& 10.39 & 37.46& 39.02& 91.86&131.80&160,350,360&78,7.8,16       &3,4,3   &31.6\pm2.2&35.0\pm1.0\\
         &    &    &       &      &      &      &      &450,800,850&3.8,0.6,1.01    &4,4,1   &          &          \\
NGC 4321 &SABb&15.2& 10.25 & 26.00& 32.28& 68.37&128.40&160,850    &46,0.88         &3,1     &30.7\pm2.1&30.5\pm1.2\\
NGC 4536 &SABb&14.9& 10.15 & 30.26& 22.49& 44.51& 54.39&850        &0.42            &1       &39.1\pm3.2&35.1\pm1.6\\
NGC 4631 &SBcd& 7.7& 10.08 & 85.40& 98.78&160.10&269.00&180,450,450&121,25.06,30.7  &5,6,1   &35.0\pm2.7&30.7\pm0.7\\
         &    &    &       &      &      &      &      &450,850,850&18,5.253,5.73   &5,6,1   &          &          \\
         &    &    &       &      &      &      &      &850,870    &1.89,3.78       &5,7     &          &          \\
NGC 5195 &SB0 & 7.7&  9.35 & 15.22& 10.85& 31.33& 12.34&850        &0.26            &1       &33.7\pm2.5&35.7\pm1.5\\
NGC 5713 &SABb&26.7& 10.54 & 22.10& 17.23& 37.28& 34.77&180,450,800&16,0.889,0.102  &5,8,8   &36.8\pm2.9&38.4\pm1.2\\
         &    &    &       &      &      &      &      &850,850,850&0.359,0.57,0.43 &9,1,5   &          &          \\
NGC 5866 &S0-a&15.3&  9.65 &  5.26&  6.66& 16.98& 16.53&180,450,850&10.5,0.79,0.14  &5,1,1   &28.3\pm1.9&31.9\pm0.8\\
NGC 6946 &SABc& 5.9& 10.08 &129.80&177.90&290.70&498.40&60,60,100  &165,115.5,338   &10,11,10&32.5\pm2.4&31.7\pm0.7\\
         &    &    &       &      &      &      &      &160,200,200&450,330,743     &12,12,10&          &          \\
         &    &    &       &      &      &      &      &200,450,850&365.8,18.53,2.98&11,1,1  &          &          \\
NGC 7331 &Sb  &13.1& 10.34 & 45.00& 56.49&110.20&164.10&60,100,200 &42.9,120,243    &10,10,10&31.4\pm2.2&28.9\pm0.8\\
         &    &    &       &      &      &      &      &450,850    &20.56,2.11      &1,1     &          &          \\
NGC 7552 &Sab &21.4& 10.86 & 77.37& 45.40&102.90& 86.65&850        &0.8             &1       &40.9\pm3.4&36.4\pm1.6\\
      \noalign{\smallskip}
      \hline
   \end{array}
   $$
{\scriptsize Columns: (3) distance in Mpc; (4) infrared luminosity in $\Lsun$ as described in Table~\ref{table:slugs450}; (5) \& (7) \IRAS flux densities in Jy; (6) \& (8) \Spitzer flux densities in Jy; (9) \& (10) Flux densities in Jy and references for the additional far-infrared and submillimeter data: 
[1] Dale et al. (\cite{Dale05}),
[2] Hunt et al. (\cite{Hunt05}),
[3] Stark et al. (\cite{Starck89}),
[4] Eales et al. (\cite{Eales89}),
[5] Bendo et al. (\cite{Bendo02}),
[6] Stevens et al. (\cite{Stevens05}),
[7] Dumke et al. (\cite{Dumke04}),
[8] Chini et al. (\cite{Chini95}),
[9] Dunne et al. (\cite{Dunne00})
[10] Alton et al. (\cite{Alton98}),
[11] Tuffs \& Gabriel (\cite{Tuffs03}),
[12] Engargiola (\cite{Engargiola91});
(12) effective temperature and $1\sigma$ uncertainties in K derived from the 60 and 100\um \IRAS flux densities; (13) effective temperature and $1\sigma$ uncertainties in K derived from all data points.}
\end{table*}

\subsection{The IR luminosity}
\label{sect:lir}
 The infrared luminosity is estimated from the bolometric luminosity of the best-fitting blackbody modified by a $\lambda^{-1.3}$ emissivity function. As a consequence, it is a measure of the IR emission of the  big dust grains in thermal equilibrium and it excludes the emission from smaller dust grains stochastically heated such as the polycyclic aromatic hydrocarbons. Distances for $H_0\!=\!75\kms$ are taken from, in decreasing order of precedence: the Catalog of Neighboring Galaxies (Karachentsev et al. \cite{Karachentsev04}), the Revised Bright Galaxy Sample (Sanders et al. \cite{RBGS}), the Surface Brightness Fluctuation survey (Tonry et al. \cite{Tonry01}), the Nearby Galaxies Catalog (Tully \cite{Tully88}) and the NASA/IPAC Extragalactic Database (NED) whose redshifts were corrected for the Virgo inflow. \IRAS flux densities are taken from, by order of precedence: the RBGS, the Large Optical Galaxies Catalog (Rice et al. \cite{Rice88}), the Point Source Catalogue with redshift (PSCz, Saunders et al. \cite{PSCz}), the Point Source Catalog (PSC, Beichman et al. \cite{PSC}) and the Faint Source Catalog (FSC, Moshir et al. \cite{FSC}).

\subsection{The star-forming size}

A well-known tracer of star formation, for which a wealth of large and high-resolution catalogs are available, and that is unaffected by dust extinction is the radio continumm. {\it Globally}, the luminosity of non-AGN galaxies at 1.4\,GHz correlates with the FIR luminosity over 4 orders of magnitudes (Condon \cite{Condon92}, Yun et al. \cite{Yun01}). {\it Spatially}, Chanial et al. (in prep.) show that the sizes of non-AGN galaxies inferred from radio continuum observations correlate tightly with the size inferred from FIR and CO maps, even for infrared luminous galaxies, in the CO case.

In this paper, the star-forming size is estimated (Chanial et al. in prep) as
\begin{equation}
\label{eq:radio-ir}
\Rir = (0.86\pm0.05)~\Rrc,
\end{equation}
\noindent
where $\Rrc$ is the HWHM of the deconvolved RC emission (see Sect. 3). The HWHM is chosen to be along the major axis, to account for the galaxy inclination at a first order.

\subsection{Infrared surface brightness}

We define the observed infrared surface brightness by the formula
\begin{equation}
\label{eq:sir}
\Sir = \frac{\Lir}{2\Air}.
\end{equation}

 The factor $\frac{1}{2}$ has been introduced to provide a better estimate of the IR emission from the effective area $\Air$, where \Rir is the observed HWHM maximum of the emission along the major-axis, because for axisymmetric gaussian profiles (face-on galaxies case), the half-light radius $r(\frac{1}{2})$=$\sqrt{2\ln2} \sigma$ is equal to the HWHM.

\section{Sample definitions}
\label{sect:samples}
Soifer \& Neugebauer (\cite{Soifer91}) showed that the complete 100\um flux-limited subsample of a complete 60\um flux-limited sample has a colder average dust temperature, which implies that the $\Lir$--\Td relation is biased by the wavelength at which a sample is selected (see also Blain et al. \cite{Blain04}).
 Furthermore, cold ultraluminous infrared galaxies falling off this relation have been discovered in the 170\um FIRBACK survey (Chapman et al. \cite{Chapman02}) and in submillimeter surveys (Chapman et al. \cite{Chapman05}).

To test whether the infrared compactness-temperature relation is subject to such a bias, we performed our analysis on two samples of non-AGN galaxies selected at 60 and 170$\um$.

\subsection{The 60\um selected sample}
\label{sect:sample60um}
\begin{figure*}
  \centering
  \resizebox{\hsize}{!}{
     \includegraphics{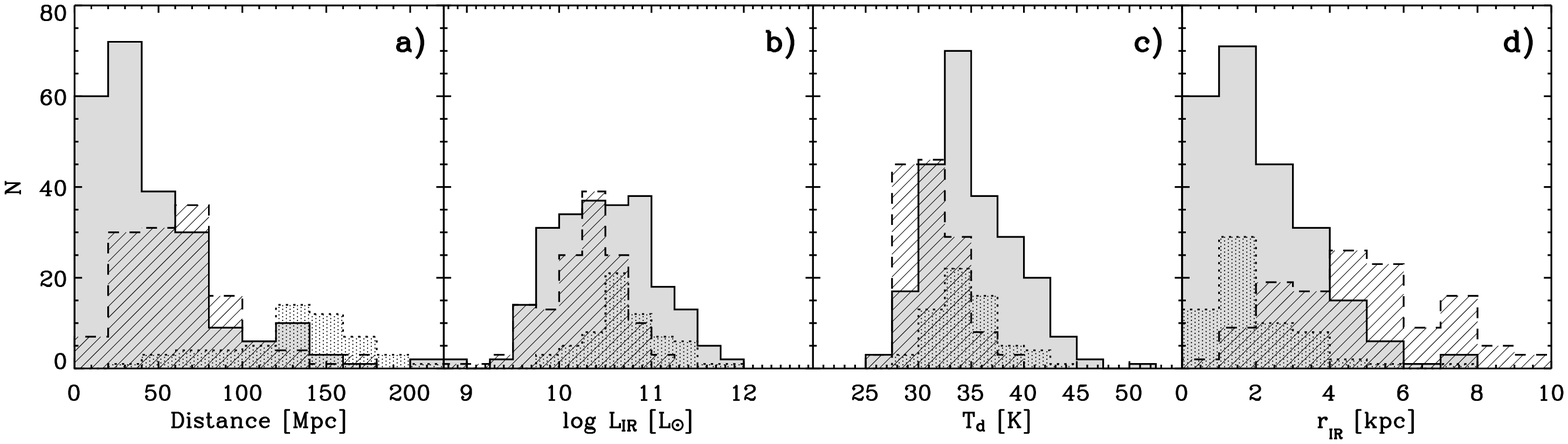}
  }
\vskip -3mm
  \resizebox{\hsize}{!}{
     \includegraphics{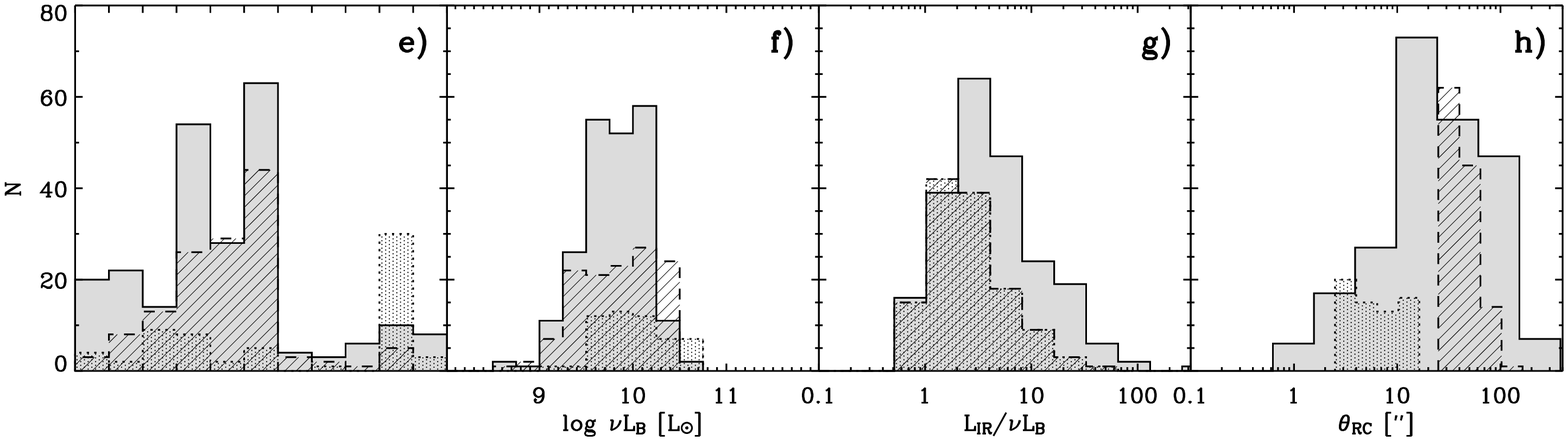}
  }
\vskip 1mm
  \caption{Distribution of {\bf a)} distance, {\bf b)} IR luminosity, {\bf c)} effective dust temperature, {\bf d)} IR physical radius, {\bf e)} morphology type, {\bf f)} B-band luminosity with $\nu$=$c/0.44\um$, {\bf g)} IR to B-band luminosity ratio and {\bf h)} FWHM of the radio continuum emission along the major axis for the 60\um sample (shaded), the 170\um sample spatially resolved in NVSS (hatched) and FIRST (dotted). The mean distance is 46.2, 62.3 and 154 Mpc respectively, the mean $\Lir$ is $10^{10.48}$, $10^{10.25}$ and $10^{10.72}\Lsun$, the mean $\Td$ is 35.01, 31.36 and 34.64\,K, the mean $\Rir$ is 2.10, 5.10 and 2.04 kpc and the mean B-band luminosity is $10^{9.80}$, $10^{9.82}$ and $10^{10.01}\Lsun$. Morphology type and B-band luminosity are taken from the HYPERLEDA database.}
  \label{fig:hist-samples}
\end{figure*}

This sample has been obtained by Chanial et al. (in prep.) by matching the RBGS, which is complete for extragalactic sources with $f_\nu(60\um)$$>$5.24\,Jy and Galactic latitudes $|b|$$>$$5\degr$, with the two RC follow-ups around 1.4\,GHz (Condon et al. \cite{Condon90}, \cite{Condon96}) of the 60\um flux-limited \IRAS Bright Galaxy Samples (Soifer et al. \cite{BGS1a}, \cite{BGS1b} and Sanders et al. \cite{BGS2}).  For each \IRAS source, one or more maps have been observed with an angular resolution ranging from 1\farcs5 to 60\arcsec. Deconvolved major-axis FWHM are provided by both datasets and are derived from two-dimensional gaussian fits to the maps.

 Starting from this initial list, we applied several editing steps to ensure the  reliability of the size estimates (cf. Wang \& Helou \cite{Wang92}, Meurer et al. \cite{Meurer97}).

\noindent
1. We retained only spatially resolved sources with angular sizes no lesser than half the radio beam FWHM.

\noindent
2. Because at higher angular resolutions, extended emission may be missed, we retained only the radio sources that contribute to more than 2/3 of the total radio continuum flux, ensuring that the angular size is representative of the whole galaxy. This step also ensures that no more than one radio counterpart is retained.

\noindent
3. For most of the \IRAS sources, no more than one RC map goes through the two steps above  and thus no more than one angular size estimate was deemed reliable. For the other sources, we adopted the mean value of the angular sizes.

\noindent
4. Contamination by an AGN has been dealt with very conservatively by excluding galaxies satistying any of the following optical, infrared and radio criteria:
(i)  position  closer than 30\arcsec\ to an AGN galaxy (including LINERs) from the extensive catalogue by V\'eron-Cetty \& V\'eron (\cite{Veron06}) or classification in NED as AGN,
(ii)  $f_{\nu}(25\um)/f_{\nu}(60\um)$$>$0.2 (de Grijp et al. \cite{deGrijp85})
and (iii) low IR-to-radio ratio $q$$<$1.94 (see Yun et al. \cite{Yun01}).

The sample resulting from this selection process contains 232 sources. They are mostly IR-dominated spirals or interacting galaxies and their global properties are summarized in Fig.~\ref{fig:hist-samples}.

\subsection{The 170\um selected sample}
The second sample has been extracted from the ISOPHOT Serendipity Sky Survey (Stickel et al \cite{ISOSSS}), which is made of 1927 sources detected at 170\um by the C200 ISOPHOT detector during the slews between pointed observations. The ISOSSS covers 15\% of the sky. A systematic inspection of the DSS (Digitized Sky Survey) snapshots of the sources and examining the FSC and the PSC lead us to discard 52 star identifications and 20 possible star associations, one likely planetary nebula, one extragalactic \Hii region, one radio source not associated with the ISOPHOT source and 7 entries that may be contaminated by a nearby radio source. We also excluded two offcenter duplicate entries of NGC\,7331 and a source contaminated by cirrus.
Before cross-correlating the ISOSSS catalog with radio catalogs, we made sure that the optical candidates taken from the HYPERLEDA database within 2\arcmin\ of the ISOSSS position had an astrometry with one arcsecond accuracy. This work resulted in the removal of 38 duplicates, the addition of 241 new sources (including 160 stars), and 515 new positions, 426 of which were taken from the 2MASS Extended Source Catalog (Jarrett et al. \cite{Jarrett00}) the remaining ones were determined on IR POSS-2 plates, which absolute astrometry were corrected using the USNO-B1 and GSC2.2 star catalogs.

Then, we cross-correlated the optical candidates with two complementary 1.4\,GHz RC surveys: the NVSS catalog (Condon et al. \cite{NVSS}) and the FIRST catalog (April 2003 release, Becker et al. \cite{FIRST}). The former has a 45\arcsec\  resolution beam and covers 82\% of the sky while the latter has a higher resolution (5\arcsec) but a more limited sky coverage (22\%). Finally, after cross-identification with the NED database, AGN candidates were removed by applying the same methodology as adopted for the 60\um sample. At this point, the sample contains 899 remaining sources.
To avoid undersampling effects, we only retained sources whose NVSS counterpart has an angular major axis 22\farcs5$<$$\Thetarc$$<$120\arcsec\  or whose FIRST counterpart has an angular major axis 2\farcs5$<$$\Thetarc$$<$15\arcsec. NVSS and FIRST major axis $\theta$ is derived from a two-dimensional gaussian fit, so that sources with a complex geometry such as spirals with dominant \Hii regions or closely interacting pairs are unlikely to be fitted. We attempted to exclude them by rejecting those with a radio-optical position offset greater than $\frac{1}{3}\Thetarc$. 
We also discarded the FIRST sources that did not make up 75\% of the total radio flux density, assumed to be the NVSS one. 

 In the final tally, the 170\um sample is made of 198 galaxies with reliable angular size information, of which 134 are taken from the NVSS catalog and the other 64 from the FIRST catalog. Like the 60\um sample, the two subsamples are mostly made of IR-dominated spirals and interacting galaxies. Their global properties are listed in the online Tables~4~and~5 and summarized in Fig.~\ref{fig:hist-samples}.

\begin{figure}
\centering
\resizebox{\hsize}{!}{
  \includegraphics{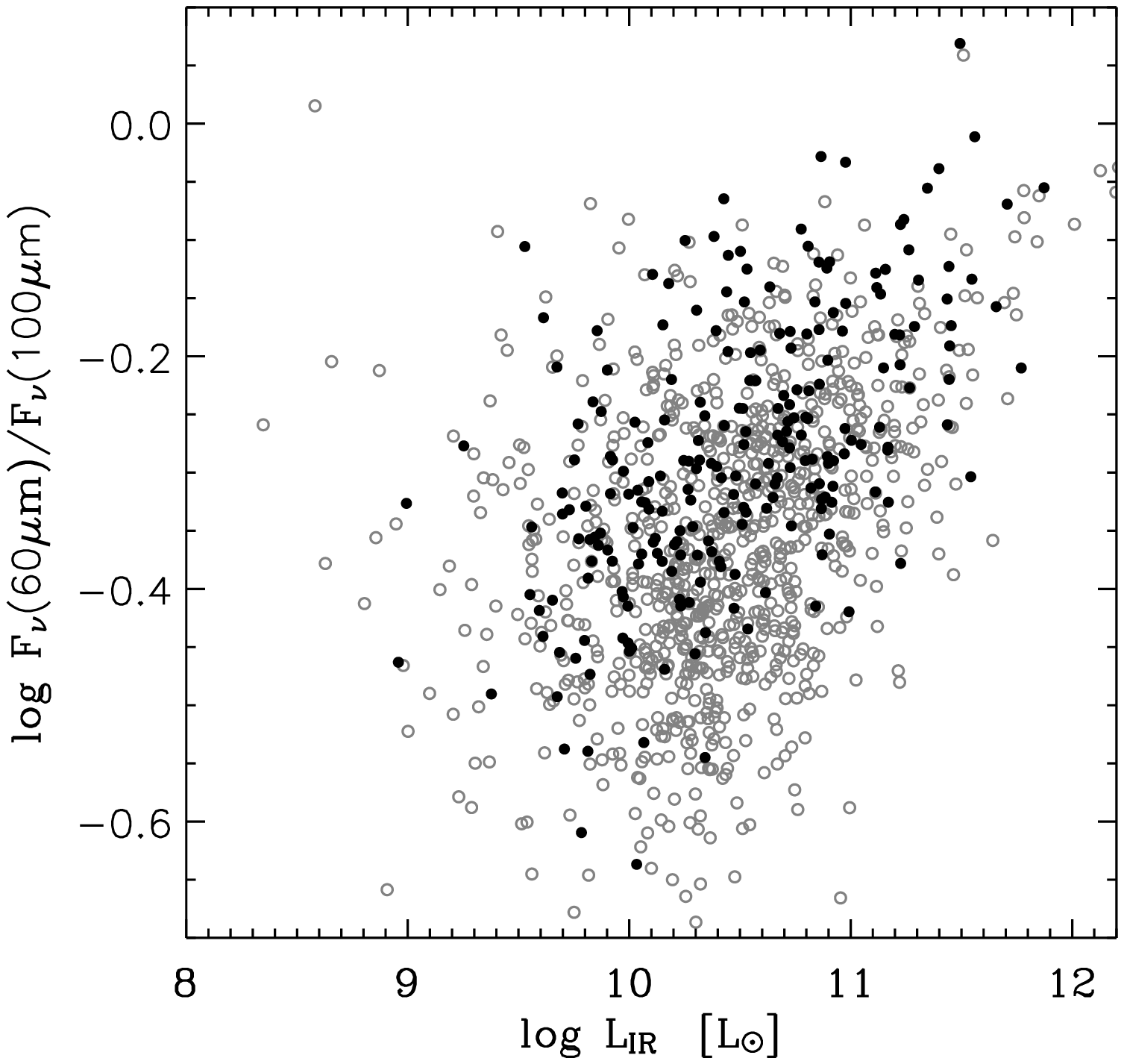}
}
\caption{The rest-frame 60-to-100\um \IRAS flux density ratio vs the infrared luminosity. The galaxies from the 60\um (170$\um$) sample are represented by filled (open) circles. Are also plotted the sources from the 170\um sample for which no reliable size estimate is available.}
\label{fig:lir-60-100}
\end{figure}

\begin{figure}
\centering
\resizebox{\hsize}{!}{
  \includegraphics{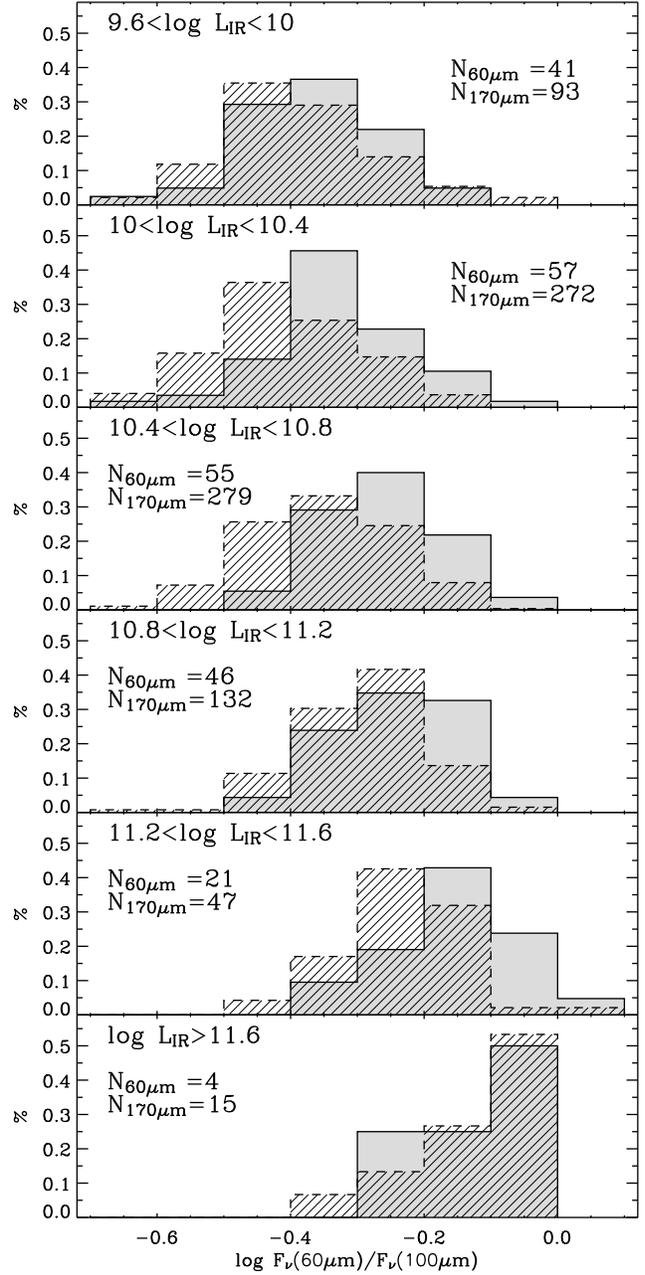}
}
\caption{Distribution of the rest-frame 60-to-100\um \IRAS flux density ratio for 6 IR luminosity bins spanning $10^{9.6}$ to $10^{12}\Lsun$. The shaded (hatched) histogram relates to the 60\um (170$\um$) sample. The 170\um sample includes the sources for which no reliable size estimate is available.}
\label{fig:hist-60-100}
\end{figure}

\begin{figure*}
\centering
\resizebox{\hsize}{!}{
  \includegraphics{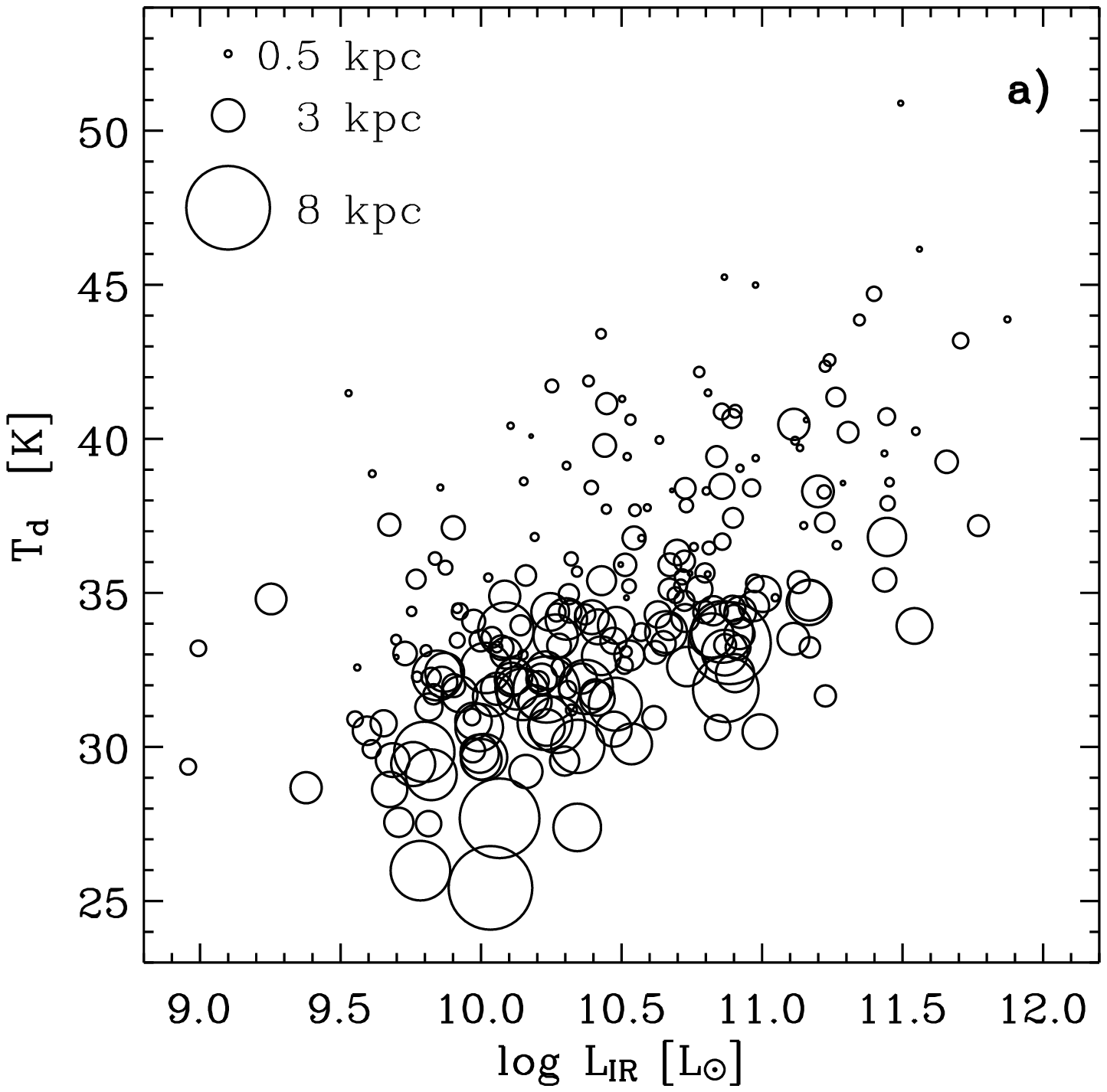}
  \includegraphics{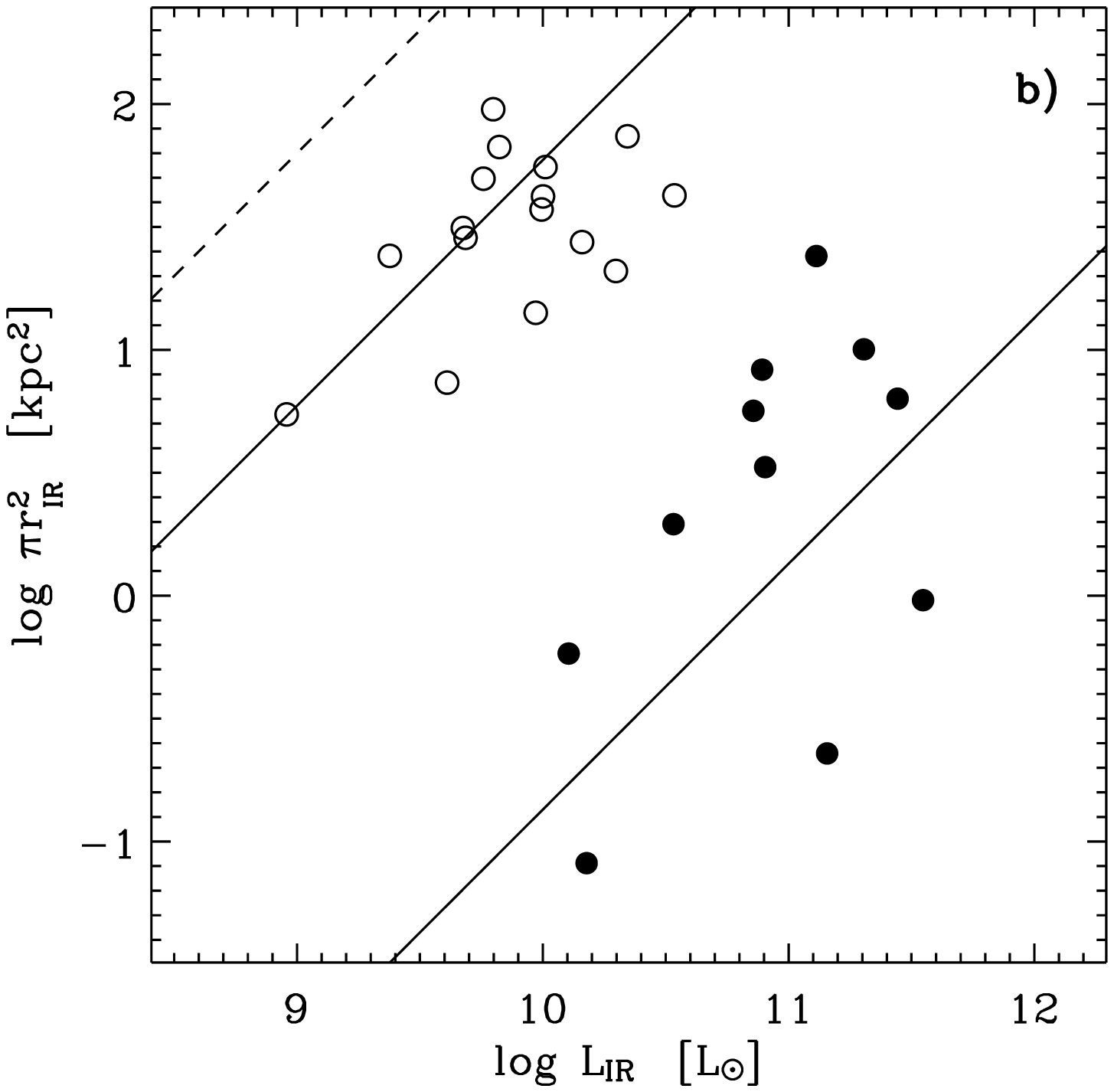}
}
\caption{{\bf a)} The \Td vs \Lir diagram for the 60\um sample. The radius of the circles is the IR radius, assumed to be proportional to the linear radio continuum HWHM along the major axis. {\bf b)} IR luminosity vs the effective IR area $\Air$\ for two temperature bins. Galaxies with 28.5$<$$\Td$$<$30.5\K are plotted with open circles and the ones with 40$<$$\Td$$<$41\K with filled circles. The solid lines corresponds to the empirical correlation given by Eq.~\ref{eq:correl-data} for the mean \Td of the two subsets. We note that the empirical relation does not exactly bisect the low-temperature subsample because of the RBGS surface brightness detection threshold (dashed line, from Wang \& Helou \cite{Wang92}) which affect the completeness of the low-temperature subsample.}
\label{fig:lir-t}
\end{figure*}

\section{Luminosity-temperature relation}
\label{sect:lt}

The response of the large dust grains to the heating radiation field has been studied through the $\Lir$--$R$(60/100) diagram. For 60\um flux-limited samples, it has been shown that the dust gets warmer as the FIR luminosity increases (Smith et al. \cite{Smith87}, Rowan-Robinson et al. \cite{RR87}, Soifer et al. \cite{BGS1b}, Soifer \& Neugebauer \cite{Soifer91}). This trend  is also followed by the 170\um sample, but more weakly.

It has been noted that the $\Lir$--$R(60/100)$ relation is affected by a large intrinsic dispersion, as shown in Fig.~\ref{fig:lir-60-100} for our 60 and 170\um samples, which is attributed to the scatter of the gas content (Soifer et al. \cite{BGS1a}, Sanders et al. \cite{Sanders91}). 
 Chapman et al. (\cite{Chapman03}) carried out a phenomenological study of this dispersion in the \IRAS 1.2 Jy survey and showed that it is key to understanding the populations of flux-limited surveys.

 The systematic effect of the sample selection on the $\Lir$--$R$(60/100) relation, which prompted us to analyse two samples selected at different wavelengths (Sect.~\ref{sect:samples}), is confirmed in Fig.~\ref{fig:hist-60-100}, in which the two samples are compared in 6 IR luminosity bins. For each bin, the 170\um sample is colder that the 60\um sample, the difference being most striking for \Lir between 10$^{10}$ and 10$^{11.6}\Lsun$.

\section{Compactness-temperature relation}

To better understand star formation processes on a global scale, we introduced an additional parameter to the $\Lir$--\Td analysis, the size of the star-forming region \Rir derived from the FWHM of the RC profiles according to Eq.~\ref{eq:radio-ir}. This new information is shown in the $\Lir$--\Td diagram (Fig.~\ref{fig:lir-t}.a) by representing the galaxies from the 60\um sample with circles of radius proportional to $\Rir$.
A systematic effect is apparent: at fixed $\Lir$, $R$(60/100) increases as the physical size decreases, reflecting the qualitative fact that the dust temperature increases as the average grain-star distance decreases.

More quantitatively, we determined the plane
\begin{equation}
\log R(60/100)=a\ +\ b\log \frac{\Lir}{\Lsun}\ +\ c\log\frac{\Rir}{\rm pc}
\end{equation}
that best matches the observations in the three dimensional parameter space. The coefficients ($a,b,c$) and their associated errors were derived by using bootstrap samples of the 60\um dataset and the resulting empirical relation is
\begin{displaymath}
\log R(60/100)=-0.663\pm0.095
\end{displaymath}
\begin{equation}
\label{eq:correl-data3}
\,\,\,+(0.092\pm0.007) \, \left[\log\left(\frac{\Lir}{\Lsun}\right)
 - (2.09\pm0.24) \, \log\left(\frac{\Rir}{\rm pc}\right)\right].
\end{equation}

\begin{figure}
  \centering
  \resizebox{\hsize}{!}{
     \includegraphics{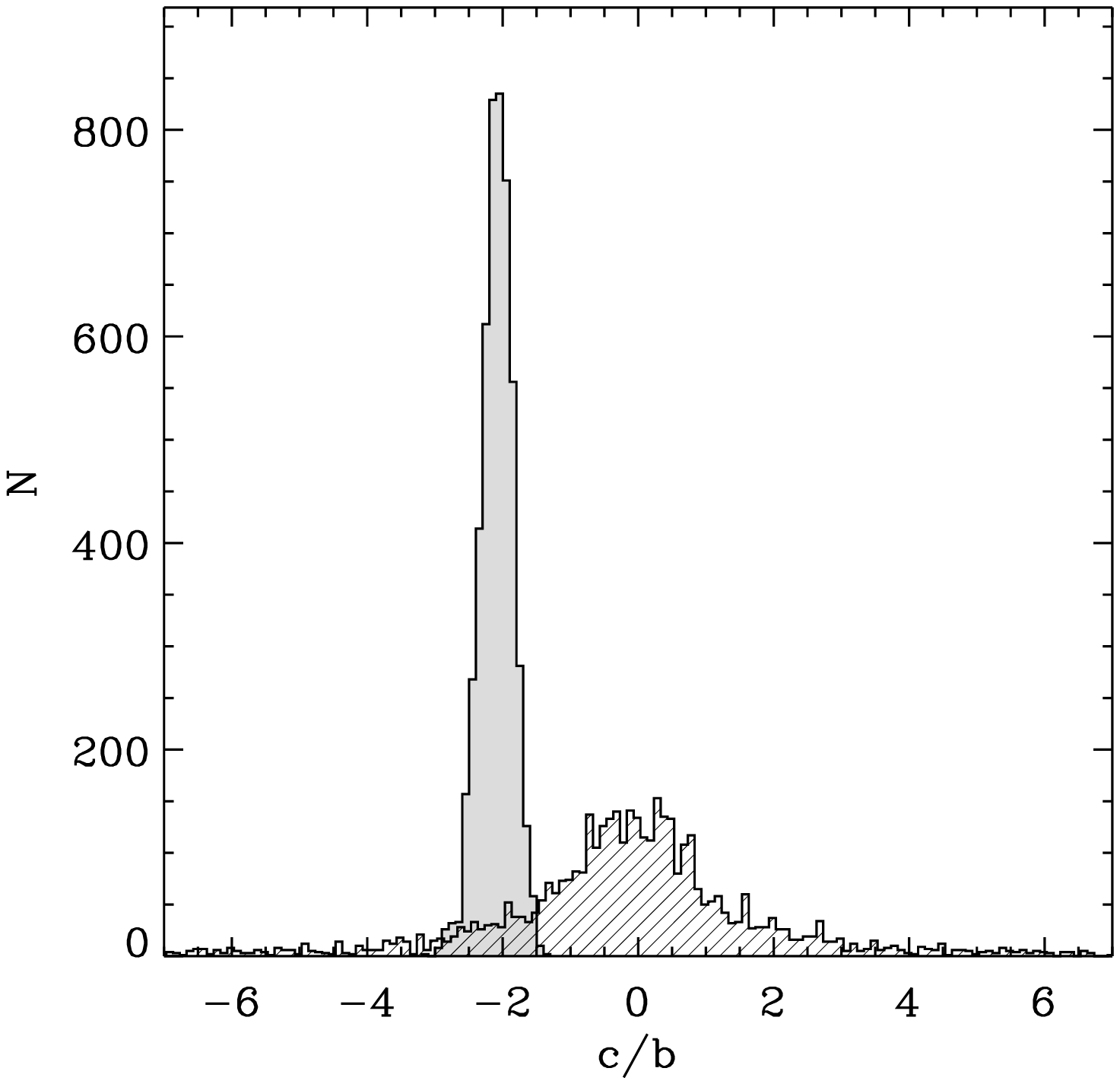}
  }
  \caption{Distribution of the values $c/b$ resulting from the regression fits of the bootstrap samples to the linear form $\log R(60/100)=a+b\log \Lir/\Lsun+c\log\Rir/{\rm pc}$. For the shaded histogram, the real observed values $R(60/100)$, $\Lir$, and \Td of the 60\um sample have been used whereas in the hatched histogram, the $R(60/100)$ values have been randomly permutated. The boostrap estimation of $c/b$ only converges for the real observed sample, to the value -2.09$\pm$0.24, which is consistent with $R$(60/100) being a function of $\Lir/\Rir^2$.}
  \label{fig:bootstrap}
\end{figure}

We note that because the covariates \Rir and \Lir scale as the distance of the galaxy and as its square, the value \hbox{$c/b$=-2} may be unduly favoured by the $\chi^2$ minimization. To check that this is not the case, we ran the bootstrap estimator on a fake 60\um sample, for which the values of \Lir and \Rir  are unchanged, but for which the values of $R$(60/100) have been randomly permutated, so that the  variate is uncorrelated with the  covariates. 
In the resulting sample, the bootstrap estimation of $c/b$ does not converge and we obtained a sample mean of $<$$c/b$$>$=-194.3 and a sample standard deviation of 891.2. In fact, its distribution is similar to a Cauchy (or Lorentz) distribution, for which moments do not exist. We note that such a similarity for the uncorrelated sample is not surprizing, because the distribution of the ratio of two independent normal distributions is precisely a Cauchy distribution. The clear difference between the distributions of the $c/b$ values for the real and the fake 60\um sample (Fig.~\ref{fig:bootstrap}) is a strong indication that the global parameters $R$(60/100), \Lir and \Rir are indeed correlated.

 The resulting value of $c/b$ is equal to -2 within its uncertainties, which implies that $R$(60/100) scales as $\Lir/\Rirtwo$ or that for a given fixed temperature, the IR luminosity linearly scales as the IR area (Fig.~\ref{fig:lir-t}.b). This finding is non-trivial and suggests that the disk geometry comes into play. Assuming a value of -2, we performed a linear regression (bisector method)

\begin{displaymath}
\log R(60/100)=-0.783\pm0.019
\end{displaymath}
\begin{equation}
\label{eq:correl-data2}
\,\,\,+(0.123\pm0.005) \, \left[\log\left(\frac{\Lir}{\Lsun}\right)
 - 2 \, \log\left(\frac{\Rir}{\rm pc}\right)\right]
\end{equation}
that can be written as
\begin{equation}
\label{eq:correl-data}
\Td = (23.5\pm0.3) \,\, \left(\frac{\Sir}{\Lsunpc}\right)^{\,0.052\pm0.002} \K,
\end{equation}
where \Td and \Sir are the effective dust temperature and the infrared surface brightness defined in Sect~\ref{sect:params}.

For the resolved 170\um sample, the relation is
\begin{equation}
\label{eq:correl-data170}
\Td= (22.9\pm0.4) \,\, \left(\frac{\Sir}{\Lsunpc}\right)^{\,0.057\pm0.004} \K,
\end{equation}
which is consistent with the equation obtained for the 60\um sample.

\section{Modelling}
Two idealized scenarios were considered to interpret the physical processes behind 
the empirical relation Eq.~\ref{eq:correl-data}: the dust being distributed in a single shell of radius equal to the observed star-forming radius $\Rir$ or in molecular clouds optically thin at far-infrared wavelengths.

\subsection{Single dust shell}
\label{sect:sds}

In this case, we assume that all the dust is distributed in an isothermal shell of radius $R$ around a point-source starburst. Lehnert \& Heckman (\cite{Lehnert96}) have proposed that this model is in agreement with the $\Sir$--\Td relation.

The energy absorbed and emitted by a grain of size $a$ and of emissivity $Q_{\rm abs}(\nu)$ is
\begin{eqnarray}
E_{\rm abs}& = & \pi a^2\int_0^{+\infty}Q_{\rm abs}(\nu)F_\nu(\nu)\ \dnu \label{eq:Eabs} \\
E_{\rm em} & = & 4\pi a^2\int_0^{+\infty}Q_{\rm abs}(\nu)\pi B_\nu(\nu,\Td)\ \dnu. \label{eq:Eem}
\end{eqnarray}

Assuming that the emissivity $Q_{\rm abs}(\nu) \simeq 1$ in the UV-optical regime and $Q_{\rm abs}(\nu) \simeq Q_{\rm abs}(\nu_0)\ (\nu/\nu_0)^\beta$ in the infrared regime, equations Eq.~\ref{eq:Eabs} and Eq.~\ref{eq:Eem} yield
\begin{eqnarray}
E_{\rm abs}/\pi a^2& = & \int_0^{+\infty} F_\nu(\nu)\ \dnu \label{eq:Eabs2}\\
E_{\rm em}/\pi a^2 & = & \frac{8\pi h}{c^2}\left(\frac{k\Td}{h}\right)^{4+\beta}\frac{Q_{\rm abs}(\nu_0)}{\nu_0^\beta}\int_0^{+\infty} \frac{x^{3+\beta}}{{\rm e}^x-1}\dx \label{eq:Eem2}
\end{eqnarray}
where $\nu_0$ is a reference frequency, $\beta$ is the dust emissivity index, and $x=h\nu/k\Td$ in the expanded Planck function.

Assuming that the dust shell is optically thick, the energy radiated by the central starburst is totally reemitted in the infrared and the right-hand part of equation Eq.~\ref{eq:Eabs2} becomes
\begin{equation}
E_{\rm abs}/\pi a^2 = \frac{\Lir}{4\pi R^2} = \frac{1}{4} \Sir. 
\end{equation}

The right-hand part of equation Eq.~\ref{eq:Eem2} can also by rewritten as
\begin{equation}
E_{\rm em}/\pi a^2 = \frac{4\sigma_\beta\ Q_{\rm abs}(\nu_0)}{\nu_0^\beta}\Td^{4+\beta} \quad\quad \beta>-3,
\end{equation}
\noindent
by using the equality (Fikhtengol'ts \cite{Fikhtengol47})
\begin{displaymath}
\int_0^\infty \frac{x^{k-1}}{{\rm e}^{ax} -1}~\dx = \frac{1}{a^k}\Gamma(k)~\zeta(k) \quad \quad k>1, \Re e~a>0,
\end{displaymath}
\noindent
where \hbox{$\Gamma(z)\!=\!\int_0^\infty t^{z-1}~{\rm e}^{-t}~\dt$} is the Gamma function and \hbox{$\zeta(z)\!=\!\sum_{n=1}^\infty 1/n^z$} is the Riemann's Zeta function, and by setting
\begin{equation}
\sigma_\beta=\frac{2\pi~k^{4+\beta}}{c^2~h^{3+\beta}}~\Gamma(4+\beta)~\zeta(4+\beta), \label{eq:stefan-boltzmann-generalized}
\end{equation}
\noindent
for which $\sigma_{\beta=0}$ is equal to the Stefan-Boltzmann constant that relates the bolometric luminosity of a blackbody to its temperature.

At the thermal equilibrium, $E_{\rm abs}=E_{\rm em}$, and the dust temperature is related to the IR surface brightness by
\begin{equation}
\label{eq:td-sir-sds}
\Td^{4+\beta}=\frac{\nu_0^\beta}{16\sigma_{\beta}Q_{\rm abs}(\nu_0)}\ \Sir.
\end{equation}

By taking $\beta\!=\!1.3$ (Sect.~\ref{sect:dust-temperature}) and the standard value $Q_{\rm abs}(125\um)\!=\!7.5\times10^{-4}$ (Hildebrand \cite{Hildebrand83}, cf. reviews by Hughes et al. \cite{Hughes97} and Alton et al. \cite{Alton04}), the numerical relation between \Td and \Sir in the single dust shell model is
\begin{equation}
\Td = 7.6\,\,\left(\frac{\Sir}{\Lsunpc}\right)^{\,0.19} \K. \label{eq:correl-sds}
\end{equation}
\noindent
It is plotted in figure Fig.~\ref{fig:sir-t} and labelled as (1a). It departs significantly from the observational data. 

To check that the difference is not due to the crudeness of the assumed dust model and to compare the work by Lehnert \& Heckman (\cite{Lehnert96}), we also used the more realistic dust model by D\'esert et al. (\cite{Desert90}). This dust model is calibrated on solar interstellar medium abundances and includes the stochastic heating of PAHs. Although it assumes an isotropic radiation field, it can be used for the single shell geometry, as long as we input the radiative energy density of the dust shell,  $u_\nu=\Sir/4c$.
The single shell model with the D\'esert et al. (\cite{Desert90}) dust model and with a heating source scaling as a O5 star is plotted in Fig.~\ref{fig:sir-t} and labelled as (1b). It shows a steep relation similar to equation Eq.~\ref{eq:correl-sds}, which is not in accordance with our samples.
 
As a result, the geometry itself of the single dust shell model is not satisfactory, unless drastic changes in the dust composition occur along the $\Sir$ sequence.
The apparent agreement between the Lehnert \& Heckman (\cite{Lehnert96}) smaller sample and the single dust shell model is likely explained by the fact that the star-forming sizes were estimated from H$_\alpha$ maps uncorrected for dust attenuation. Because extinction preferentially affects high compactness regions, their maps likely missed the central nuclei of the most compact starbursts, leading to an overestimation of the star-forming sizes and an underestimation of the infrared surface brightness.

\subsection{UV-optical-thick \& FIR-thin molecular clouds}
\label{sect:thinmodel}
 We assume that the infrared emission is from a disk of radius $R$ and that the dust giving raise to this emission is isothermal. Several dust configurations could yield an isothermal dust population such as thin shells around young star clusters or dust in cirrus exposed to a uniform interstellar radiation field.

We also assume that every dust grain radiates as a blackbody modified by a $\lambda^{-\beta}$ emissivity function. In the case in which the medium is transparent in the far-infrared but opaque in the optical-UV, the luminosity \Lir radiated by the dust is proportional to the total mass of dust $M_{\rm d}$ and more specifically (Hildebrand \cite{Hildebrand83})

\begin{equation}
F_\nu(\nu) = \kappa(\nu) \, M_{\rm d} \, \frac{B_\nu(\nu,\Td)}{D^2},
\end{equation}
where $F_\nu$ is the flux density, $\kappa(\nu)$ is the absorption mass coefficient assumed to be equal to $\kappa(\nu_0)~(\nu/\nu_0)^\beta$  and $D$ is the distance. The integration of this relation over the frequencies leads to

\begin{equation}
\Lir = 4\pi~D^2~\int_0^{\infty}F_\nu(\nu)~\dnu = 4~\sigma_\beta \frac{\kappa(\nu_0)}{\nu_0^\beta}~M_{\rm d}~\Td^{4+\beta},
\end{equation}
where $\sigma_\beta$ is the ``generalized'' Stefan-Boltzmann constant which we introduced in Eq.~\ref{eq:stefan-boltzmann-generalized}.

Setting the dust-to-gas mass ratio \mbox{$\eta_{\rm d}\!=\!M_{\rm d}/M_{\rm gas}$}, the total luminosity can be related to the total gas mass $M_{\rm gas}$ by
\begin{equation}
\label{eq:bb}
\Lir = 4~\sigma_\beta \frac{\kappa(\nu_0)}{\nu_0^\beta}~\eta_{\rm d}M_{\rm gas}~\Td^{4+\beta}.
\end{equation}

The size of the star-forming region is introduced in our analysis through the Schmidt law that non-linearly relates the star formation rate to the gas mass surface density with a power index that reliably departs from unity. This law has been most accurately determined by Kennicutt (\cite{Kennicutt98}) as

\begin{equation}
\label{eq:schmidt}
\frac{\Sigma_{\rm SFR}}{\Msun\,{\rm yr^{-1} \, kpc^{-2}}} = 2.5\,10^{-4} \, \left(\frac{\Sigma_{\rm gas}}{\Msun\,{\rm pc^{-2}}}\right)^{1.4},
\end{equation}

\noindent
where $\Sigma_{\rm gas}$ is the gas mass surface density inside the radius $R$.

Kennicutt (\cite{Kennicutt98}) derived the star formation rate from the total infrared luminosity

\begin{equation}
\label{eq:sfr-lir}
 \frac{\rm SFR}{\Msun\,{\rm yr^{-1}}} = \gamma\ \frac{\Lir}{5.8 \, 10^9\Lsun}
\end{equation}
\noindent
 by assuming that the dust absorbs and reprocesses all the intrinsic star light (factor $\gamma$=1).

Because Eqs. \ref{eq:bb} and \ref{eq:sfr-lir} are linear, both of their sides can be divided by  $\pi R^2$ and expressed in terms of surface densities and surface brightness.
By substituting $\Sigma_{\rm SFR}\!=\!{\rm SFR}/\pi R^2$ from Eq.~\ref{eq:sfr-lir} in Eq.~\ref{eq:schmidt}, we can derive the gas mass surface density as a function of the infrared surface brightness. Then, $\Sigma_{\rm gas}\!=\!M_{\rm gas}/\pi R^2$ can be eliminated from Eq.~\ref{eq:bb} and we obtain the following relation between \Td and \Sir

\begin{equation}
\label{eq:td-sir-literal}
\Td^{4+\beta}=6.27\ 10^{-5} \ \frac{\nu_0^\beta}{\sigma_{\beta}\ \kappa(\nu_0)\ \eta_{\rm d}\ \gamma^{1/1.4}}
\left(\frac{\Sir}{\Lsunpc}\right)^{0.4/1.4}.
\end{equation}

Assuming that the interstellar medium is optically thick, $\gamma$=1 and by adopting standard values of the parameters involved in our model,
$\kappa(125\um)\!=\!1.9~{\rm m}^2~{\rm kg}^{-1}$ (Hildebrand \cite{Hildebrand83}),
$\eta_{\rm d}\!=\!1/350$ (see Sanders et al. \cite{Sanders91}, Bendo et al. \cite{Bendo03})
and $\beta\!=\!1.3$, Eq.~\ref{eq:td-sir-literal} gives

\begin{equation}
\label{eq:correl-thin}
\Td = (22.9\pm0.9)\,\,\left(\frac{\Sir}{\Lsunpc}\right)^{\,0.054\pm0.013} \K.
\end{equation}

\begin{figure*}
\centering
\resizebox{\hsize}{!}{\includegraphics{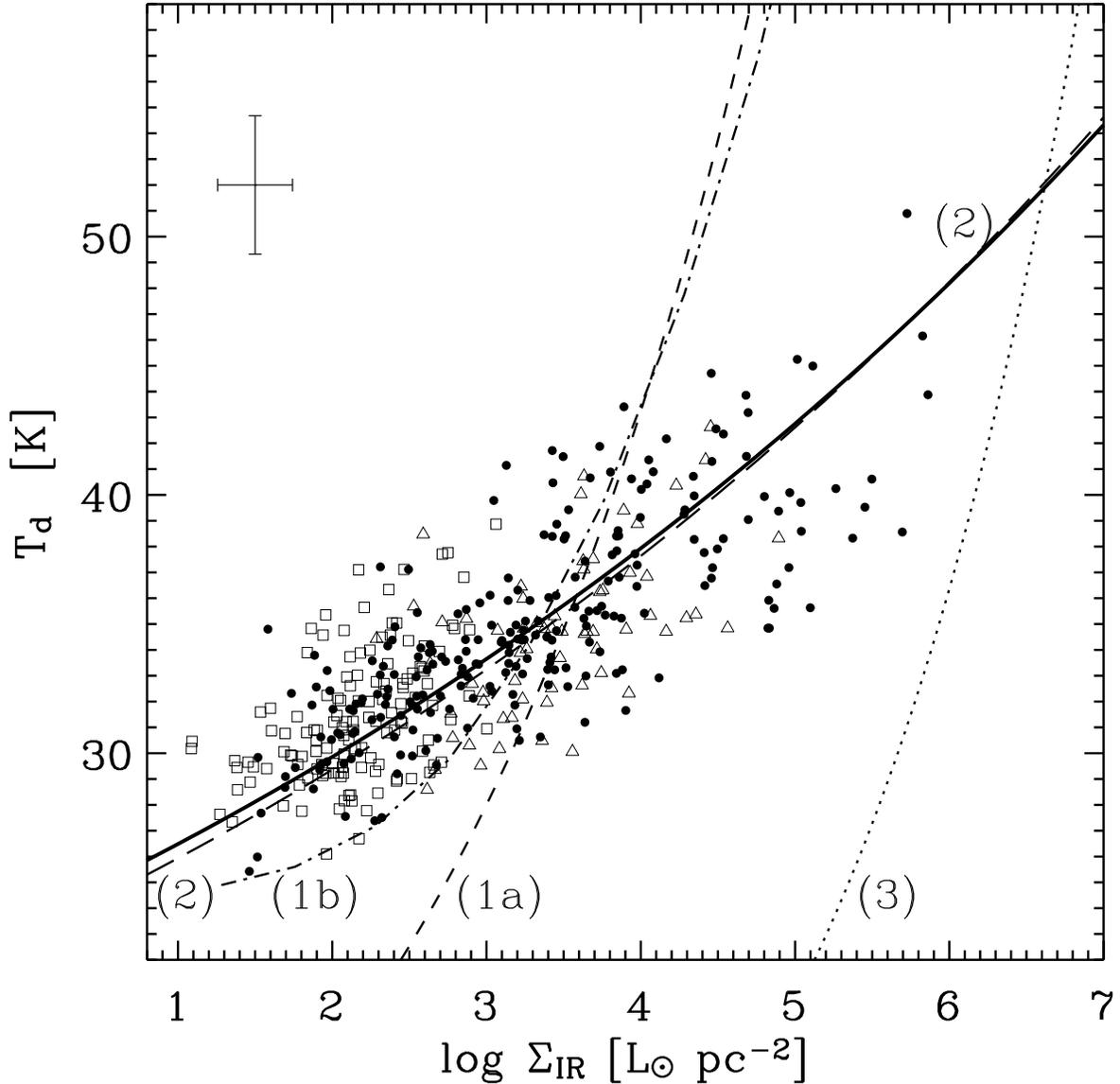}}
\caption{The effective dust temperature $\Td$ vs. the IR surface brightness $\Sir$ for the 60\um sample (filled circles) and the 170\um sample (open squares and triangles for the sources with reliable NVSS and FIRST angular sizes respectively). The solid thick line shows the empirical relation Eq.~\ref{eq:correl-data}. The dashed line (1a) represents (1a) the single dust shell model with an analytical dust model (Sect.~\ref{sect:sds}, Eq.~\ref{eq:correl-sds}), and the dot-dashed line (1b) the single dust shell model coupled with the D\'esert et al. (\cite{Desert90}) dust model. The long-dashed line (2) is the model involving the Schmidt law (Sect.~\ref{sect:thinmodel}, Eq.~\ref{eq:correl-thin}) and the dotted line (3) is the expected relation for a blackbody (Eq.~\ref{eq:correl-bb}). The average uncertainties in \Td and \Sir are shown based on errors of 10 per cent for $f_\nu(60\um)$, 15 per cent for $f_\nu(100\um)$ and 20 per cent for $\Rir$.}
\label{fig:sir-t}
\end{figure*}

This equation is plotted in Fig.~\ref{fig:sir-t} and labelled as (2). It is in very close agreement over 5 orders of magnitude with the empirical relation Eq.~\ref{eq:correl-data} obtained from the 60\um sample of galaxies and is in accord with the 170\um sample, though we should bear in mind that the actual position of the theoretical relation could be translated along the $\Td$-axis because of the uncertainties in $\kappa(\nu_0)$ and $\eta_{\rm d}$.

\section{Discussion}

\subsection{Selection biases}

Sources with $10^{10}$$<$$\Lir$$<$$10^{10.5}\Lsun$ are on average colder in the 170\um NVSS-resolved subsample ($<$$\Td$$>$=30.5$\K$) than in the 60\um sample (34.0$\K$). The difference is not the result of the additional selection criteria on the NVSS sources, because their average effective dust temperature is comparable to the whole 170\um sample in this luminosity bin. The temperature difference arises from the wavelength selection that affects the $\Lir$--\Td diagram, as discussed in Sect.~\ref{sect:lt}. On the contrary, the $\Sir$--\Td relation is followed by both the  60\um and 170\um selected samples and we conclude that the latter relation is not sensitive to the selection wavelength, and reflects more fundamental physical processes in play in quiescent and starburst galaxies.

 Another selection bias that could affect the $\Sir$--\Td relation is the Malmquist bias  involved in the determination of the Tully-Fisher relation for spiral galaxies (Tully \& Fisher \cite{Tully77}) and the Fundamental Plane of elliptical galaxies (Dressler et al. \cite{Dressler87}, Djorgovski \& Davis, \cite{Djorgovski87}). This bias would underpopulate galaxies with small physical sizes, which would lead to an underestimation of the effective dust temperature of the most compact galaxies. This effect may be present in the 170\um samples for which the angular size is censored and to a lesser extent in the 60\um sample. However, out of the 75 \IRAS sources observed in Condon et al. (\cite{Condon90}, \cite{Condon96}) with an angular resolution higher than or equal to 1\farcs8, only 9 unresolved sources would have passed the selection criteria of the 60\um sample described in Sect.~\ref{sect:sample60um}. 

Our samples are selected at infrared wavelengths and the $\Sir$--\Td relation may not stand for galaxies fainter in the infrared or with a low metallicity. However, it is unclear whether we should expect  higher or lower dust temperatures for these types of galaxies because two factors may be in competition.
On one hand, the calibration of the SFR-to-\Lir ratio in Eq.~\ref{eq:sfr-lir} assumes an optically thick dust model. The stellar emission is strongly attenuated in the galaxies of our samples, as shown by the IR to B-band luminosity ratio in Fig.~\ref{fig:hist-samples}.g and this assumption is justified for our samples. However, galaxies for which only a fraction of the intrinsic young stellar emission is attenuated by the dust would have higher SFR-to-\Lir ratios ($\gamma$$>$1) which would give in Eq.~\ref{eq:td-sir-literal} lower dust temperatures for a given IR surface brightness.
On the other hand, for example, Wilson et al. (\cite{Wilson91}) found that the SFR estimated by the IR luminosity was greater than the one estimated by the (extinction-corrected) H$_\alpha$ emission line in the coolest region of the spiral galaxy M\,33 and attributed this excess to the presence of interstellar cirrus (Helou \cite{Helou86}). They can represent a non-negligeable fraction (up to 50\%-70\% or more, Lonsdale-Persson \& Helou \cite{Lonsdale87}, Bell \cite{Bell03}) of the total IR emission in quiescent galaxies. An increase of the interstellar cirrus contribution would decrease the SFR-to-\Lir ratio ($\gamma$$<$1) that would in turn increase the dust temperature for a given IR surface brightness.
However, it should be noted that the most quiescent galaxies in our samples do not  depart from the $\Sir$--\Td relation so it is possible that the two factors compensate themselves or have second order effects.

It should also be noted that the dust in this model is in thermal equilibrium and as a consequence, the stochastic heating of small dust grains is not taken into account. However, with our choice of observables, this fact is mitigated by the adopted definition of \Lir (Sect.~\ref{sect:lir}) which also discards the mid-IR excess emission expected from the smaller dust grains.

\subsection{Emissivity index dependence}

The modelled $\Sir$--\Td relation (Sect.~\ref{sect:thinmodel}) has a weak dependence on the emissivity index $\beta$ as shown in table~\ref{table:sir-t-beta} by the modest variation of the scaling factor $a$ and the power-law index $b$ given by the relation $\Td=a\,\Sigma_{\rm IR}^{\,b}$. The modelled values of $a$ and $b$ have been calculated with the same values of the absorption mass coefficient $\kappa(\nu_0)$ and dust-to-gas ratio $\eta_{\rm d}$ as in Sect.~\ref{sect:thinmodel} ($\beta$=1.3). Because of the uncertainties associated with these two parameters, the comparison between the empirical and modelled scaling factors $a$ is not insightful, as already noted. More noteworthy, the empirical and modelled power-law indices $b$ do not depart from each other as $\beta$ varies, which makes the agreement of the model with the observations not sensitive to our initial choice of emissivity index.

\begin{table}
   \caption{Dependance of the $\Sir$--\Td relation on the emissivity index $\beta$. The effective dust temperature is given by $\Td=a\ \Sigma_{\rm IR}^{\,b}$ with \Sir being in unit of$\Lsunpc$. The empirical relation is fitted to the 60\um sample. The model is described in Sect.~\ref{sect:thinmodel} and assumes an absorption mass coefficient $\kappa(125\um)\!=\!1.9~{\rm m}^2~{\rm kg}^{-1}$ and a constant dust-to-gas mass ratio
$\eta_{\rm d}\!=\!1/350$.\label{table:sir-t-beta}}
   $$
\begin{array}{l cccc}
      \hline
      \hline
      \noalign{\smallskip}
      & \multicolumn{2}{c}{\rm Observation} & \multicolumn{2}{c}{\rm Model} \\
\beta & a & b & a & b\\
\hline
      \noalign{\smallskip}
1   &24.32&0.0549& 22.77&0.0571\\
1.5 &22.96&0.0503& 22.94&0.0519\\
2   &21.68&0.0466& 22.88&0.0476\\
      \noalign{\smallskip}
      \hline
\end{array}
$$
\end{table}

\subsection{The $\Lir$--\Td degeneracy}
With conservative errors of 10 and 15 per cent for the \IRAS flux densities at 60 and 100$\um$, the statistical standard deviation of the effective dust temperature of the 60\um dataset is expected to be 2.8\K. It is compatible with the sample standard deviation, which is lower than 3.2\K along the $\Sir$--$\Td$ sequence. It implies that the introduction of the size parameter has mostly disentangled the degeneracy of the $\Lir$--\Td diagram down to current observational precision. The FIR all-sky survey by the satellite {\it Akari} will be able to address further this issue.

\subsection{A starburst temperature limit?}

One can wonder what part of the $\Sir$-$\Td$ diagram would be populated by ultra-compact starbursts not resolved or too faint to be included in our samples and for which the assumption of FIR-thin molecular clouds may be  incorrect. As the distribution of IR emission becomes more compact, the FIR opacity would increase until the source becomes a blackbody, for which the IR surface brightness is related to the temperature  by $\Td=(\Sir/4\sigma_{\beta=0})^{1/4}$, or numerically by
\begin{equation}
\label{eq:correl-bb}
\Td =1.15 \,\,\left(\frac{\Sir}{\Lsunpc}\right)^{\,0.25}\K
\end{equation}
\noindent
Ultra-compact starbursts would lay in Fig.~\ref{fig:sir-t} on the left-hand side of the steep Eq.~\ref{eq:correl-bb} and on the upper side of Eq.~\ref{eq:correl-thin}. In this scenario, the effective dust temperature would not be constrained.
However, the FIR-opacity of the most active starbursts is still a matter of debate. Another scenario proposed by Klaas et al. (\cite{Klaas01}) for ultra-luminous galaxies driven by star-formation, is that such galaxies are still mostly FIR-thin and as a consequence, they would still follow the relation given by Eq.~\ref{eq:correl-thin} until they reach the empirical starburst intensity limit by Meurer et al. (\cite{Meurer97}) of $2.0\times10^5\Lsunpc$ (with a factor of 3 uncertainty). With these assumptions, the effective temperature would then be limited to around 44$\K$ for 90\% or more starbursts.

\section{Conclusions}

We constructed two well-defined local datasets selected at 60\um and 170$\um$, which are made of 430 IR-bright, non-AGN galaxies with reliable radio continuum sizes. We used them to 
 investigate the relation between three global parameters, namely the infrared luminosity $\Lir$, the effective dust temperature \Td and the size of the star-forming region $\Rir$. We show that

1. The dispersion in the $\Lir$--\Td diagram can be explained by introducing the size of the star-forming region.

2. Infrared bright non-AGN galaxies form a plane in the ($\Lir,\Td,\Rir$) space akin to the fundamental planes of the spiral and elliptical galaxies.

3. The effective dust temperature is related to $\Lir$/\Rirtwo, {\it i.e.} the IR surface brightness, by a power-law over 5 orders of magnitude. 

4. Unlike the $\Lir$--\Td relation, the $\Sir$--\Td relation does not depend on the IR wavelength used to select or detect galaxy samples.

5. The empirical relation is in agreement with a simplified model made of isothermal molecular clouds which are opaque in the optical
and transparent in the FIR and for which we assumed a constant emissivity index,
gas-to-dust ratio and mass absorption coefficient.

6. The model for which the dust is distributed in a single shell around the central starburst is ruled out.

7. Because \Td also traces the SFR per unit of gas mass, the SFR per unit of gas mass correlates with the SFR per unit area.

8. The infrared compactness turns out to be a parameter able to describe the smooth sequence ranging from quiescent to starburst galaxies in which the gas surface density, the effective dust temperature, the SFR per unit of gas mass and the SFR per unit area increase together.

Assuming that the relation holds for distant galaxies, they may be significant, since the hierarchical framework of structure formation predicts a decrease of sizes with redshift. It would induce an evolution of galaxy colors, which may statistically affect the derivation of the cosmic star formation rate from infrared galaxy number counts. Such a study will be carried on a subsequent paper. 

The unprecedented angular far-infrared resolution of ESA's {\it HERSCHEL} space observatory will allow us to further probe the infrared compactness-temperature relation. In a more distant future, high redshift galaxies will be spatially resolved by missions included in the ESA Cosmic Vision Programme such as the  Far InfraRed Mission ({\it FIRM}) or in the NASA space science roadmap such as: the Single Aperture Far--IR telescope ({\it SAFIR}), the Space IR Interferometric Telescope ({\it SPIRIT}) and the Submillimeter Probe of the Evolution of Cosmic Structures ({\it SPECS}).

\begin{acknowledgements}
P. Chanial acknowledges financial support from the National Research Council, E. Dwek for his advices and stimulating discussions and S. Madden, M. Vaccari for their inputs. The anonymous referees are thanked for their feedback.
This research has made use of the HYPERLEDA database (http://leda.univ-lyon1.fr) and the NASA/IPAC Extragalactic Database (NED) which is operated by the Jet Propulsion Laboratory, California Institute of Technology, under contract with the National Aeronautics and Space Administration. This work has also made use of photographic data obtained using The UK Schmidt Telescope.
        The UK Schmidt Telescope was operated by the Royal Observatory
        Edinburgh, with funding from the UK Science and Engineering Research
        Council, until 1988 June, and thereafter by the Anglo-Australian
        Observatory.  Original plate material is copyright (c) the Royal
        Observatory Edinburgh and the Anglo-Australian Observatory.  The
        plates were processed into the present compressed digital form with
        their permission.  The Digitized Sky Survey was produced at the Space
        Telescope Science Institute under US Government grant NAG W-2166.

\end{acknowledgements}

\end{document}